\documentclass[a4paper,12pt]{article}

\begin{document}

\title{Estimating Renyi entropies of a multiparticle system from
event-by-event fluctuations\footnote{Dedicated to Adriano Di Giacomo on
the occasion of his 70th birthday.} }

\author{A.Bialas and K. Zalewski \\ M.Smoluchowski
Institute of Physics \\Jagellonian University,
Cracow\footnote{Address: Reymonta 4, 30 059 Krakow, Poland,
e-mail: bialas@th.if.uj.edu.pl, zalewski@th.if.uj.edu.pl}
\\ and
\\ Institute of Nuclear Physics, Polish Academy of Sciences}
\maketitle

\begin{abstract}
Recent improvements in the method of estimating Renyi entropies
from measurements of coincidences between the events observed in
high energy collisions are reviewed. A new, more precise,
formulation of the method is presented and its accuracy analyzed.
\end{abstract}

 \section {Introduction}

Entropy of a system produced in high-energy collisions is an
interesting object, very useful for understanding  the physics of
the process in question. This is particularly important for the
search for quark-gluon plasma in heavy ion collisions. It is not
easy, however, to obtain information on entropy directly from data
(without additional assumptions about the properties of the
system). A window which may open such a possibility is to study
the coincidences between observed events. As was suggested in
\cite{bc}, such measurements may allow to estimate of the Renyi
entropies of the system \cite{ren} and thus, by extrapolation,
give information on its Shannon entropy. This simple idea (based
on an old suggestion by Ma \cite{ma}) is, however, difficult to
implement \cite{dl} and its accuracy hard to determine. These two
problems were studied recently in a series of papers by W.Czyz and
the present authors \cite{bcz,bczf}. In this note we compile and
summarize these results . Although no new results are presented
(all can be found in \cite{bcz} and \cite{bczf}), we feel that
such a compilation in a single place will be convenient for the
reader and may be useful for future applications.

The object of our study is the M-particle semi-inclusive distribution.
It is defined by considering a collection of events in which exactly $M$
particles were observed in a given region of the momentum space. We
shall call them $M$-particle events (independently of how many particles
were actually produced)\footnote{This terminology is often used in
experimental descriptions of multiparticle processes. The proper
technical terms are: exclusive distribution if all particles are
observed, and semi-inclusive distribution if besides a given number of
observed particles there is an unspecified number of other particles.
This should not be confused with inclusive $M$-particle distributions.}.
These events can be described by the normalized $M$ particle Wigner
function $W_M(X,K)$ with $X=X_1,...,Z_M$, $K=K^{(1)}_x,...,K^{(M)}_z$,
which we shall interpret as the M-particle phase-space distribution
\cite{wig}.

It should be emphasized that the phase-space distribution $W_M$,
describing the semi-exclusive distribution, refers only to particles
actually {\it measured} in a given experiment and in a given momentum
region. It gives no direct information about the particles which are not
registered by the detector. To discuss the phase-space density of {\it
all} produced particles, additional assumptions (e.g. of thermodynamic
equilibrium) are necessary.

At this point it is also important to realize that the phase-space
distribution of particles produced in high-energy scattering is not a
precisely defined quantity. Apart from the standard problems with the
uncertainty principle, one has to take into account that particles may
be produced at different times. In the present paper, following
\cite{ber1}, we shall consider the time-averaged distribution.

The aim of this paper is to discuss (i) how the moments of $W_M(X,K)$
can be estimated from the measured coincidences of the observed events
and, (ii) how these moments are related to Renyi entropies and thus also
to the Shannon entropy of the system.

To this end we first introduce the {\it effective} coincidence
probabilities $\hat{C}_M(l)$ of order $l$,
 related to the moments of the phase-space distribution
 by\footnote{To simplify the formulae we shall from now on omit the
index $M$ in all quantities. Since we are discussing solely $M$-particle
events, this should not lead to any confusion.} \cite{bcz}

\begin{equation}\label{e1}
\hat{C}(l)= (2\pi)^{3M(l-1)} \int d^{3M}X \int d^{3M}K [W(X,K)]^l.
\end{equation}

These quantities are interesting because, as was  shown in
\cite{bcz} (and will be explained in the next section),
for a rather large class of  phase-space densities,
 $\hat{C}(l)$ defined above can be approximated by
the measured coincidence probability $C^{exp}$ of the M-particle
events \cite{bc,bc2}
\begin{equation}\label{}
C^{exp}(l) =\frac {N_l}{N(N-1)...(N-l+1)/l!} \label{31}
\end{equation} where $N_l$ is the number of the
observed l-plets of identical events and $N$ is the total number of
events. $N(N-1)...(N-l+1)/l!$ is the total number of l-plets of
events\footnote{For l=2 formula (\ref{31}) was first suggested, in a
different context, by Ma \cite{ma}. See also \cite{bcw}.}.
One sees that the measurement of $C^{exp}_l$
reduces to the count of the number of coincidences between the observed
events.

As the next step we investigate the relation between $\hat{C}(l)$,
as defined by (\ref{e1}), and the coincidence probabilities of
 the states of the system, $C(l)$, given by
\begin{equation}\label{}
C(l)\equiv \sum_i[P_i]^l= Tr[\rho^l] \label{i2}
\end{equation}
where the sum runs over all states of the system,
$P_i$ is the probability of a state $i$ to occur and $\rho$ is the
density matrix of the system. The second part of this equality is
obvious in the representation where the density matrix is diagonal.
Since
the trace of a matrix is independent of the representation,  the
result is generally valid.

 $C(l)$ defines the Renyi entropy of order $l$,  $H(l)$, by the formula
\begin{equation}\label{}
H(l)= \frac 1{1-l}\log C(l) \label{i1}
\end{equation}
 and thus opens a window
to the true entropy of the system. Indeed, as is well known and easy to
show
\begin{equation}\label{i3}
S=\lim_{l\rightarrow 1} H(l)
\end{equation}
where $S=-\sum_iP_i\log
P_i=-Tr[\rho \log \rho]$ is the Shannon entropy.

Unfortunately, since measurement of coincidences can only provide
information on $H(l)$ for integer $l\geq 2$ and, in practice, only for
$l=2,3$ and perhaps $l=4$, the extrapolation procedure is rather
uncertain \cite{zy}. However, since for $l\geq 1$ \cite{beck},
\begin{equation}\label{i4}
S \geq H(l)\geq H(l+1) \end{equation}
 Renyi entropies provide an {\it exact lower limit} for $S$, a
quantity very important for understanding the properties of the
quark-gluon plasma \cite{plasma}.

We show that $C(l)$ and $\hat{C}(l)$ are equal to each other in the
limit of infinite size of the system. Also the finite volume corrections
are studied and shown to fall with inverse square of the smallest
(linear) size. When combined with the previous result, one obtains a
reliable method to measure, with a good control of error, the Renyi
entropies of the system.

In the next section an Ansatz for the particle phase-space distribution
is introduced. In Section 3 the corresponding formulae for the effective
coincidence probabilities $\hat{C}(l)$ are written down and the optimal
binning procedure is obtained by comparing them with $C^{exp}(l)$. In
Section 4 the true coincidence probabilities $C(l)$ are analyzed in the
same framework and relation between $\hat{C}(l)$ and $C(l)$ is
explained. Our conclusions and outlook are given in the last section.

\section {The phase-space distribution}

To proceed we consider a rather general form of the
 phase-space distribution
\begin{equation}\label{36a}
W(X,K)= \frac1{(L_xL_yL_z)^M}G[X/L] e^{-v(K)} \end{equation} with
$X/L\equiv (X_1-\bar{X}_1)/L_x,...,(Z_M-\bar{Z}_M)/L_z$,
 $K= K_1,...,K_{M}$. The function $G$
satisfies  the normalization conditions
\begin{eqnarray}\label{37}
\int d^{3M}u G(u)&=& 1\;\;\rightarrow\;\;\int dX G(X/L) =
(L_xL_yL_z)^M ;\nonumber\\
 \int d^{3M}u u_i G(u)&=&
0\;\;\rightarrow\;\; <X_i,Y_i,Z_i>=
\bar{X}_i,\bar{Y}_i,\bar{Z}_i;\nonumber\\
 \int d^{3M}u(u_i)^2
G(u) &=& 1\;\;\rightarrow\;\; <(X_i-\bar{X}_i)^2,...> =L_x^2,...
\end{eqnarray}

The first condition insures that $e^{-v(K)}$ is the properly normalized
(multidimensional) momentum distribution\footnote{By definition
 of the phase-space distribution,  the momentum distribution
is given by $\int dX W(X,K)$.}, the second defines the central values of the particle
distribution in configuration space and the third defines $L_x,L_y,L_z$
as  the root mean square sizes of the distribution in configuration
space. Both sizes and central positions may depend on the particle
momenta\footnote{They may be also different for different kinds of
particles.}. The form of the function $G$ describes the shape of the
multiparticle distribution in configuration space.

The form (\ref{36a}) for the time-averaged phase space density is
satisfied in a  variety of models \cite{mod}. It is obviously valid
for models which assume thermal equilibrium. It can also incorporate
expansion of the system, provided $\bar{X}_i's$ depend on $K_i's$ (the
Hubble-like expansion is obtained for $\bar{X}_i \sim K_i$). It is
general enough to incorporate any multiparticle momentum distribution.

In our further discussion we shall restrict somewhat this general form
by taking the function $G(u)$ as a Gaussian:
\begin{equation}\label{}
G(u) = \frac1{(2\pi)^{3M/2}} e^{-\sum_{m=1}^M\sum_\alpha [u_{m\alpha}]^2/2}
\label{gauss}
\end{equation}
where $m$ labels the particles and $\alpha$ labels the space directions.
This restriction can be avoided at the cost of some complications of the
algebra. Since the exact shape of the particle emission region is not
well determined and since, moreover, (\ref{gauss}) is not in obvious
disagreement with the data from quantum interference, we shall stick to
it.

\section{Moments of phase-space distribution and
experimental coincidence probabilities}

Using the Ansatz for the phase space density given by (\ref{36a}) and
(\ref{gauss}), we discuss in this section the relation of {\it
experimental} coincidence probabilities (\ref{31}) to the {\it
effective} coincidence probabilities, determined by the moments of the
phase-space semi-inclusive densities  (\ref{e1}).

To discuss $C^{exp}(l)$ we have to face the problem of discretization.
The point is that the measured events are characterized by particle
momenta which are continuous variables. Therefore, the definition
(\ref{31}) is not directly applicable: a binning is necessary. Once
discretized, the identical events can be defined as those which have the
same population of the predefined bins and thus counting of coincidences
becomes straightforward\footnote{A detailed description of this
procedure was given in \cite{dl} and applied in \cite{kit}.}. The
counting of identical events obviously depends on the binning, however,
so the procedure is ambiguous \cite{bc,dl,bc2,bcw}. In order to obtain
a viable estimate of $\hat{C}(l)$, we thus have to select the binning in
such a way that the result of (\ref{31}) is as close as possible to that
given by (\ref{e1}).

Let us denote the $3M$-dimensional momentum bins by $j = 1,\ldots ,J$
and their volumes by $\omega_j$. As the first step we express the
measured l-fold coincidences (\ref{31}) by the $3M$-dimensional
distribution of momenta

\begin{equation}\label{o1}
  e^{-v(\textbf{K})} = \int\!\!dX\;W(\textbf{K},\textbf{X})
\end{equation}
and the binning $\omega_j$. This is clearly possible because the
observed coincidences depend only on the momentum distribution and
on binning.  The relevant formulae are derived in \cite{bcz}.

Next, we
consider   $\hat{C}(l)$, defined in (\ref{e1}).
Using (\ref{36a}) and (\ref{gauss}), a formula for $\hat{C}(l)$
can be written down in form of an integral
\begin{equation}\label{38}
\hat{C}(l)= \int  d^3K_1...d^3K_M  \hat{\Omega}(K_1,...,K_M;l)
\end{equation} where
\begin{equation}
\hat{\Omega}(K_1,...,K_M;l)= \frac{(2\pi)^{3M(l-1)/2}}{l^{3M/2}}
\frac{ e^{-lv(K_1,...,K_M)}  } {(L_xL_yL_z)^{(l-1)M}}. \label{381}
\end{equation} The derivation can be found in \cite{bcz}.

The integral (\ref{38}) can be of course replaced by a sum over all
bins $\omega_j$.

The last step is to compare the formulae for $C^{exp}(l)$ and
$\hat{C}(l)$. When this is done, one observes that if the bin sizes are

\begin{equation}\label{omo}
  \omega_j = \prod_{m=1}^M
\left[\frac{(2\pi)^{3(l-1)/2}}{l^{3/2}(L_xL_yL_z)_j^{(m)}}\right]
=
\prod_{m=1}^M\prod_\alpha
\left[\frac{(2\pi)^{(l-1)/2}}{l^{1/2}(L_\alpha)_j^{(m)}}\right]
\end{equation}
where $\alpha=x,y,z$, we obtain

\begin{eqnarray}
\hat{C}(l) =
C^{exp}(l)\frac{\sum_j\frac1{\omega_j}\int_{\omega_j}dK
e^{-lv(\vec{K})}}
{\sum_j[\frac1{\omega_j}\int_{\omega_j}dKe^{-v(\vec{K})}]^l}. \label{finl}
\end{eqnarray}
Note that the product $L_xL_yL_z$ is related to the volume of the system
in configuration space.

Eq. (\ref{omo}) is very general and can be applied to an entirely
arbitrary discretization procedure. In the simple (but most useful in
practice) case when $L_x,L_y,L_z$ do not depend on $\vec{K}$,
the components $K_x,K_y,K_z$ can be  divided into bins of
constant lengths $\Delta_x,\Delta_y,\Delta_z$. Then the condition for
the size of the bin is

\begin{equation} \label{rel1}
\Delta_x\Delta_y\Delta_z=
\left(\frac{(2\pi)^{l-1}}{l}\right)^{3/2}\frac1{L_xL_yL_z}.
\end{equation}
and $\omega_j=[\Delta_x\Delta_y\Delta_z]^M$.

Equations (\ref{finl}) and (\ref{rel1}) define  the method of
estimating the effective coincidence probabilities $\hat{C}(l)$ from the
observed coincidence probabilities $C^{exp}$.

The formula (\ref{finl}) can be rewritten in a somewhat more intuitive
form
\begin{eqnarray}
\hat{C}_M(l) =
C^{exp}_M(l)\frac{\sum_{bins} <e^{-lv(\vec{K})}>}
{\sum_{bins}<e^{-v(\vec{K})}>^l}. \label{rel}
\end{eqnarray}
which explicitely shows that in the limit when  bins are so small
that the momentum distribution inside each bin can be treated as a
constant, $\hat{C}(l)= C^{exp}(l)$. This implies that, as discussed in
detail in \cite{bcz}, the accuracy of the method improves for large
volume $(L_xL_yL_z)$ of the system. It also shows that the method of
estimating the Renyi entropies proposed in \cite{bc} is only an
approximation, valid at a very large volume of the system.

For smaller systems the correction factor
\begin{equation}
\Phi\equiv\frac{\sum_{bins} <e^{-lv(\vec{K})}>}
{\sum_{bins}<e^{-v(\vec{K})}>^l}
\end{equation}
 may be estimated either
from the measured single-particle  distribution and correlation
functions\footnote{$\Phi$ depends only on momentum distribution.} or,
more precisely, by Monte Carlo simulations.

\section{Moments of phase-space distribution and Renyi entropies }

In this section we discuss the relation between the effective
coincidence probabilities $\hat{C}(l)$ (defined by the moments of the
phase-space distribution (\ref{e1})), and the coincidence probabilities
of the states of the system (defined in terms of the density matrix
(\ref{i2})). To this end we have to evaluate the trace of the $l$-th
power of the density matrix and compare it with the formula for
$\hat{C}(l)$.

The density matrix can be obtained from the phase-space distribution
(Wigner function) by the Fourier transform \cite{wig}:
\begin{eqnarray}
\rho_M(p;p')\equiv \rho(p_1,...,p_M;p_1',...,p_M')= \int dX e^{i q
X} W(X,K)\equiv\nonumber\\
{}\equiv \int d^3X_1...d^3X_M e^{i[q_1X_1+...+q_MX_M]}
W(X_1,...,X_M,K_1,...K_M)= \nonumber\\=e^{-v(K_1,...,K_M)}
e^{-\frac12 \sum_{m=1}^M \sum_\alpha L^2_\alpha q_{m\alpha}^2 + i
\sum_{m=1}^M \sum_\alpha q_{m\alpha}\bar{X}_{m\alpha}(K) }
 \label{r1}
\end{eqnarray}
where $q=p-p'$ and $K=(p+p')/2$, and where we have explicitely used the
Gaussian form (\ref{gauss}) of  $G(u)$.

It is seen from this formula that $\rho(p;p')$ may be diagonal in $p,p'$
only if $W[X,K]$ does not depend on $X$, a condition which can be
 realized only if the volume of the sytem extends to
infinity.

The non-diagonal nature of the density matrix is in fact the fundamental
reason for all complications.

When (\ref{r1}) is introduced into (\ref{i2}) we  obtain $C(l)$ in form of
a multidimensional integral. In \cite{bczf} the first two terms in the
asymptotic
form of this integral at large $L_x,L_y,L_z$ were investigated.
The results are summarized in the formula
\begin{equation}
C(l)= \int  d^3K_1...d^3K_M  \Omega(K_1,...,K_M;l)  \label{yrr1}
\end{equation}
where
 \begin{equation}
\Omega(K_1,...,K_M;l)=
\frac{\hat{\Omega}(K_1,...,K_M;l)}{\Theta(K_1,...,K_M;l)}. \label{w12}
\end{equation}
$\hat{\Omega}(K_1,...,K_M;l)$ is defined in (\ref{381}) and
\begin{equation}
\Theta(K_1,...,K_M;l)=Det\left[1+\sum_{s=1}a_s T^s\right]\label{w2}
\end{equation} with
\begin{equation}
a_s= \frac1{2^{2s}}\frac{(l-1)!}{(2s+1)!(l-2s-1)!} \label{w33}
\end{equation} and $T$ is the $3M \times 3M$ matrix
\begin{equation}
T_{m\alpha,n\beta}=\frac1{L_\alpha}V_{m\alpha,n\beta}\frac1{L_\beta}.
\label{tt} \end{equation} where \begin{equation}
V_{m\alpha}=\partial_{m\alpha} v(K_1,...,K_M)\;;\;\;
V_{m\alpha,n\beta}=\partial_{m\alpha}\partial_{n\beta}
  v(K_1,...,K_M).\label{vdef}
\end{equation}
The indices $m,n=1,...,M$ denote particles and $\alpha,\beta= x,y,z$ denote
directions.

Note that the sum over $s$ is finite, because all $a_s$ vanish for $2s >
l-1$. In particular, for $l=2$ all $a_i=0$ and we simply have
$\Omega(K,;2)=\hat{\Omega}(K;2)$.

Comparing (\ref{yrr1}) with (\ref{38}) one sees that for $l\geq 3$
$\hat{C}(l)$ and $C(l)$ differ only by the factor
$\Theta(K_1,...,K_M;l)$ under the integral. The first observation is
that in the limit when all $L_x,L_y,L_z$ are very large, the matrix
$T_{m\alpha,n\beta}$ tends to 0 and thus the correction factor $\Theta$
approaches $1$. Consequently, the difference between $\hat{C}(l)$ and
$C(l)$ disappears. For finite size of the system, using (\ref{w12}), the
difference can be explicitely calculated from the M-particle momentum
distribution.

 \section {Discussion }

When combined together, the results reported in Sections 3 and 4,
provide a substantial improvement on the method of estimating the Renyi
entropies of a multipartcile system suggested originally in \cite{bc}.
First, the discretization procedure, necessary to give a precise meaning
to the coincidence measurement, is properly formulated. Second, the
role of the size of the system in configuration space for the accuracy
of the measurement is explained. Finally, the corrections due to the
finite size of the system are explicitely evaluated and can be used to
improve the precision of the method. In effect we obtain a practical and
reliable method of determining the Renyi entropies of the multiparticle
systems and thus also the lower limit for its Shannon entropy.

Several comments are in order.

(i) One sees from (\ref{omo}) that the optimal size of the bin does not
depend on the average position of the particles at freezeout
$\bar{X}(K)$. One also sees from (\ref{vdef}) - (\ref{tt}) that the
correction factor $\Theta$ does not depend on it. This implies that the
momentum-position correlations induced by the $K$-dependence of
$\bar{X}$ do not influence significantly the measurement of Renyi
entropies.

(ii) It is also seen from (\ref{omo}) that only the volume of the bin
$\omega_{j_m}=(\Delta_x\Delta_y\Delta_z)_{j_m}$, but not its shape,
matters in the determination of the optimal discretization. One can use
this freedom to improve the accuracy of the measurement by taking bins
large in the directions with weak momentum dependence and small in the
direction where the momentum dependence is significant.

(iii) Our analysis can be applied to any part of the momentum space. This
is important for two reasons. First, for large systems, when the optimal
size of the bins is small, a reliable measurement of coincidencies in
full momentum space may require a prohibitively large statistics. Thus
restriction to a small part of phase space may be necessary. Second, it
allows to measure the local entropy density in momentum space
(integrated over all configuration space). In case of strong
momentum-position correlations, the selection of a given momentum region
can induce, however, a selection of a corresponding region in
configuration space.

(iv) The accuracy of the measurement depends on the correct estimate of
the size of the system. Information from HBT measurements allowing to
determine the parameters $L_x,L_y,L_z$ (and, hopefully, also the shape
of the emission region \cite{dan}) is therefore necessary. One should
keep in mind, however, that the interpretation of the results from
quantum interference is far from unique \cite{bk}. Thus some modelling
may be needed.

(v) It is interesting to note that the effect of the finite size of the
system  tend to cancel when the Eqs. (\ref{rel}) and
(\ref{w12}) are combined. Indeed, increasing $L$ implies
smaller bins and thus {\it decreasing} $C^{exp}(l)$. Thus also
$\hat{C}(l)$ evaluated from (\ref{rel}) {\it decreases}. But one sees
from (\ref{w12}) that then $C(l)$ {\it increases}. Thus the obtained
value of Renyi entropy is less sensitive to a change in size of the
system than Eqs. (\ref{rel}) and (\ref{w12}), taken separately, suggest.

(vi) Through this paper we have only discussed the M-particle systems
(at fixed $M$). The coincidence probabilities including all
multiplicities can obtained from the relation
\begin{eqnarray}
C(l) =
\sum_M [P(M)]^l C_M(l) \label{16}
\end{eqnarray}
where $P(M)$ is the multiplicity distribution. One sees from this
formula that at large $l$ only multiplicities close to the most probable
one contribute effectively to $C_l$.

(vii) It should be emphasized that the method we propose takes into
account all correlations between particles measured in the experiment.
This is to be contrasted with the estimate of entropy from the single
particle inclusive distribution \cite{palb} where, by definition, the
correlations between paticles are neglected (the assumption of
equilibrium is used instead).

To summarize, we have reviewed recent developments \cite{bcz,bczf} of
the method of estimating the Renyi entropies from measurement of
coincidences between events observed in high energy collisions
\cite{bc}. As discussed in detail, the accuracy of the method depends
crucially on the size of the system in configuration space. It turns out
that the original idea is strictly correct only for systems of very
large size. The finite size corrections are derived. As shown in
\cite{bczf} they are negligible for systems encountered, e.g., in heavy
ion collisions. For smaller systems they are more important but seem not
to be prohibitive even for systems of the size as small as 1 fm. One
thus obtains a new, reliable, tool for studies of the effective degrees
of freedom in multiparticle phenomena.

The proposed method does not demand any assumptions about the
thermodynamic properties of the system, in particular it does not assume
 thermodynamic equilibrium. It thus may be of particular interest for
testing the standard assumptions of the models of quark-gluon plasma.
Moreover, it can serve as a quantitative measure of the deviations from
 equilibrium.

\section*{Acknowledgments}

We have greatly profited from the active participation of Wieslaw
Czyz at the origin of this project. We would like to thank him for
allowing to use here his results prior to publication. Discussions
with Jacek Wosiek and Robi Peschanski are also acknowledged. This
investigation was partly supported by the MEiN research grant 1
P03B 045 29 (2005-2008).


\begin{thebibliography} {99}
\bibitem{bc} A.Bialas and W.Czyz, Phys. Rev. D61 (2000) 074021.
\bibitem{ren}
A.Renyi, Proc. 4-th Berkeley Symp. Math. Stat. Prob. 1960, Vol.1, Univ.
of California Press, Berkeley-Los Ageles 1961, p.547. \bibitem{ma} S.K.Ma,
Statistical Mechanics, World Scientific, Singapore 1985; S.K.Ma, J.
Stat. Phys. 26 (1981) 221.
\bibitem{dl} A.Bialas and W.Czyz, Acta Phys.
Pol. B31 (2000) 687.
\bibitem{bcz} A.Bialas, W.Czyz and
K.Zalewski, hep-ph/0506233; Acta Phys. Pol. B36 (2005) 3109
[hep-ph/0508289].
\bibitem{bczf} A. Bialas and K. Zalewski, hep-ph 0512248, to be published in
 Acta Phys.Pol. B; A.Bialas, W.Czyz and K.Zalewski, hep-ph/0512293.
\bibitem{wig} For a discussion of the physical meaning of
the Wigner function see, e.g., M.Hillery, R.F.O'Connell, M.O.Scully and
E.P.Wigner, Phys.Rept. 106 (1984) 121.
\bibitem{ber1} G.F.Bertsch, Phys. Rev.
Letters 72 (1994) 2349; 77 (1996) 789 (E); D.A.Brown,
S.Y.Panitkin and G.F.Bertsch, Phys. Rev. C62 (2000) 014904.
\bibitem{bc2} A.Bialas and
W.Czyz, Acta Phys. Pol. B31 (2000) 2803; B34 (2003) 3363.
\bibitem{bcw}
A.Bialas, W. Czyz and J.Wosiek, Acta Phys. Pol.B30 (1999) 107 .
\bibitem{zy} K.Zyczkowski, Open Sys. and Information Dyn. 10 (2003) 297.
\bibitem{beck} C.Beck and F.Schloegl, Thermodynamics of chaotic systems,
Cambridge U. Press, Cambridge (1993).
\bibitem{plasma} For a recent discussion
see, e.g. B.Muller and K.Rajagopal, hep-ph/0502174 and references
therein.
\bibitem{mod}  For a review of models see, e.g., U.A. Wiedemann
and U. Heinz, Phys. Rep. 319(1999)145; U. Heinz and B. Jacak, Ann. Rev.
Nucl.Part.Sci. 49(1999)529; R.M. Weiner, Phys. Rep. 327(2000)250;
T.Csorgo, H.I.Phys. 15 (2002)1.
\bibitem{kit} K.Fialkowski and R.Wit,
Phys.Rev. D62 (2000) 114016; NA22 coll, M. Atayan et al., Acta Phys.
Pol. B36 (2005) 2969.
 \bibitem{dan} See, D.A.Brown and P.Danielewicz, Phys. Lett. B398 (1997)
252; Phys. Rev D58 (1998) 094003; S.Y.Panitkin and D.A.Brown, Phys. Rev
C61 (1999) 021901; G.Verde et al, Phys. Rev. C65 (2002) 054609;
P.Danielewicz et al., Acta Phys. Hung. A19 (2004) nucl-th/0407022.
\bibitem{bk} See, e.g., A.Bialas and K.Zalewski, Phys. Rev. D72 (2005) 036009.
\bibitem{palb} S.Pal and S.Pratt, Phys. Lett. B578 (2004) 310.

\end{thebibliography}
\end{document}